\documentclass[12pt,preprint]{aastex}

\received{}
\revised{}
\accepted{}
\ccc{}
\cpright{}{}

\slugcomment{}
\shorttitle{M13 Subgiants}
\shortauthors{Briley, Cohen, \& Stetson}

\newcommand{\subsun}{\mbox{$_{\odot}$}}
\newcommand{\teff}{$T_{eff}$}
\newcommand{\grav}{log($g$)}
\newcommand{\kms}{km~s$^{-1}$}
\newcommand{\fe}{[Fe/H]}
\newcommand{\cabund}{[C/Fe]}
\newcommand{\ciso}{$^{12}$C/$^{13}$C}

\begin{document}

\title{Carbon Abundances of Faint Stars in M13  - Evidence for 
Two Abundance Altering Mechanisms
\altaffilmark{1}}

\author{Michael M. Briley}
\affil{Department of Physics and Astronomy, University of Wisconsin
Oshkosh, 800 Algoma Boulevard, Oshkosh, WI 54901}

\author{Judith G. Cohen}
\affil{Palomar Observatory, MS 105-24, California Institute of Technology,
Pasadena, CA 91125}

\and

\author{Peter B. Stetson}
\affil{Dominion Astrophysical Observatory, 5071 West Saanich Road,
Victoria, BC V9E 2E7, Canada}

\altaffiltext{1}{Based on observations obtained at the
W.M. Keck Observatory,  which is operated jointly by the California 
Institute of Technology, the University of California, and the
National Aeronautics and Space Administration.}

\begin{abstract}
We present an analysis of CH band strengths in Keck LRIS spectra of a 
sample of 81 stars in M13 within 2 magnitudes of the main-sequence turnoff.
The  subgiants clearly exhibit a substantial (a factor of $\sim$6) spread in \cabund.  
Moreover, the bulk of the subgiants possess C abundances larger than those
found among their more luminous counterparts.
The turnoff stars themselves are too warm for appreciable CH formation, but the
relatively small range in the observed CH band strength for stars just below
the turnoff nevertheless translates into this same spread in \cabund.
Still fainter, the sample size is small, but the same range in \cabund\ appears to
be present.
On the basis of these observations we suggest that a process external to
the present stars has resulted in a substantial star-to-star dispersion in
\cabund\ (and possibly other light elements) among all stars in M13.
In addition, the surface C abundances among the more
luminous stars have been further modified by the operation of an internal
deep-mixing mechanism during red giant branch ascent.   
The amplitude of the scatter we find in \cabund\ at all luminosities may
prove difficult to explain via accretion from intermediate mass
AGB stars as the external ``polluting'' mechanism.

\end{abstract}


\section{INTRODUCTION}

A detailed history of the thirty years of progress in our understanding
and interpretation of the compositions of globular cluster stars is far too long and
complex to offer in the present work. 
Hence we present below only the shortest of summaries.
For a more proper treatment, we recommend to the reader the more comprehensive
reviews of \citet{Kraft94} and \citet{DaCosta98}.

Contrary to the canonical picture of globular clusters as chemically homogeneous
populations of stars, significant star-to-star differences in the abundances of C,
N, and often other light elements associated with proton capture reactions, have
been found in every Galactic globular cluster studied to date.
Locked within these inhomogeneities appears to be at least
a partial history of processes at work in the early cluster environment.
However,  these patterns are also likely subject to further modification
driven by a less than well-understood mixing mechanism at work within the
stars themselves.
With this in mind, intra-cluster abundance inhomogeneities are usually
discussed in terms of these two distinct origins: internal 
post-first dredge-up  ``deep mixing'' or a process external to the present
day cluster stars (e.g., the accretion of ejecta from more massive stars).

In the case of deep mixing, material exposed to the CN (and possibly ON) cycle
regions above the H-burning shell of an evolving low-mass star is
circulated into the outer convective envelope during red giant branch (RGB) ascent
resulting in increased surface N abundances at the expense of C (and possibly O).
Observational evidence that such a process operates can be found
among low-mass metal-poor ($-2 \le$ \fe $\le -1$) field giants where
\citet{Gratton00} note lowered
C abundances, enhanced N, depleted Li, and lower \ciso ~ ratios than
expected from first dredge-up \citep*[see also][]{Charbonnel98,Keller01}.
Mixing may be even more efficient in the metal-poor ($-2 \approx$ \fe) globulars where
substantial decreases in \cabund ~ ($\approx$ 1 dex) with increasing luminosity
have been observed along the RGBs of M92 \citep{Bellman01}, M15
\citep{Trefzger83}, and NGC 6397 \citep{Briley90}.
Mixing has also been suggested as an explanation for the detections
of O underabundances and O-Na, Al-Mg anticorrelations among the bright
giants of numerous globular clusters, e.g. M5, M3, M10, M13
\citep[see the composite figure for O-Na relations assembled by][their Fig. 12,
and references therein]{Ramirez02} - trends notably absent from the field
population \citep{Gratton00}.
If proton capture reactions on Ne and Mg resulting in Na and Al are
also operating at ON-cycle temperatures \citep[][and references
therein]{Cavallo00} one could then expect the
abundance patterns observed in most cluster giants to be at least qualitatively
reproduced under the conditions of sufficiently deep mixing.

But the underlying physical process connecting the shell region with the
envelope beyond first dredge-up is not included in standard models of low
mass stellar evolution and remains poorly understood.
Some possible mechanisms explored include meridional circulation
\citep{Sweigart79}, various treatments of diffusion \citep[e.g.][]{Charbonnel95, 
Denissenkov00}, and shell flashes \citep*{Aikawa01} - see the
review of the theory of red giant branch stars
by \citet*{Salaris02} for a more through discussion.

As early as 24 years ago it was known that the subgiant branch (SGB) stars and
likely the main-sequence (MS) stars of the metal-rich (\fe $= -0.8$)
cluster 47 Tuc exhibit C and
N differences similar to those found on its RGB \citep{Hesser78}.
These results have been confirmed by \citet[and references therein]{Cannon98},
and significant C/N inhomogeneities among SGB and MS stars have since been observed
in NGC 6752 and M5 \citep*{Suntzeff91,Cohen02},
and M71 \citep{Cohen99a}.
Variations in other light elements have also been noted among the SGB/MS stars
of these very clusters: Na-CN correlated on the MS of 47 Tuc \citep{Briley96},
O-Na and Mg-Al anticorrelated among the MS stars of NGC 6752
\citep{Gratton01}, and O-Na anticorrelated in M71 SGB stars \citep{Ramirez02}.
Most recently, \citet{Ramirez03} have shown a significant dispersion in Na extending
from the M5 RGB to its MS  turnoff. 
These inhomogeneities are of course difficult to explain within the framework
of mixing theories and imply that some component of the
abundance differences is in place well before appreciable RGB ascent,
possibly embedded in the present-day stars much earlier in the cluster
history.
Indeed, the nuclei observed to vary among the SGB/MS stars, i.e., C, N, O, Na,
Al, and Mg, suggest as a source the hot bottom burning regions of intermediate mass
asymptotic giant branch (AGB) stars \citep{Ventura01}.
As put forth by \citet{Cottrell81}, the proton-exposed ejecta from such
stars may then have been accreted by and thus ``polluted'' the present day cluster
stars \citep[see][for a more thorough discussion]{Cannon98}.

It has become increasingly clear that a full description of cluster abundance
inhomogeneities requires an understanding of the extent of the contributions
of {\it both} deep internal mixing and external sources of chemical differences.
Unfortunately, the same reaction chains (proton captures on C, N, O, Ne and Mg)
appear to be associated with both mechanisms.
Perhaps the only method of disentangling the role of each process
is to compare the abundances of the less luminous cluster stars with their
more evolved counterparts.
Because of the strength of the CH band, for technical reasons this is easier to accomplish
at present for C than for the other light elements listed above.

In this regard, the globular cluster M13 provides us with a particularly interesting
target.
With an overall metallicity of \fe = $-1.51$ dex
\citep{Kraft92}, it is perhaps
one of the most  thoroughly studied ``intermediate metallicity'' clusters.
It has long been known to contain giants with a large and bimodal spread in
CN band strengths, anticorrelated with CH, whose origin lies in an
apparent C-N anticorrelation \citep[e.g.][]{Smith96}.
In addition, among its RGB stars, large\footnote{As commented by \citet{Kraft97},
the M13 RGB stars exhibit some of the largest inhomogeneities known.}
O differences are present (anticorrelated with N), as well as significant star-to-star
scatter in Na and Al \citep[Na anticorrelated with Mg and correlated with O, see][]
{Kraft97, Cavallo00}.
Such observations are suggestive of the mixing of proton exposed
material to the surfaces of some cluster members and there is evidence
of shifts towards lower O and higher Na and Al abundances among the
most luminous RGB stars ($M_V < -1.7$) \citep{Pilachowski96, Kraft97,
Cavallo00}.
Yet, at the same time, a large dispersion in these elements is present in
the data of Kraft et al. to at least within a 0.25 mag of the luminosity
function (LF)  bump at $V \approx 14.75$ \citep{Paltrinieri98}.
A large scatter in C and N abundances was also followed to a
similar luminosity by \citet{Suntzeff81}.
\citet*{Grundahl98} found scatter in Str\"{o}mgren
photometry among the RGB and SGB stars in M13 which they attributed to a
spread in CNO abundances probably extending at least to the main-sequence
turnoff.

We report here on recent observations and derived C abundances of a sample
of SGB/MS stars in the globular cluster M13 which appear to support the
hypothesis of a substantial intrinsic spread in the distribution of light elements
within the cluster which is further modified during RGB ascent.

\section{OBSERVATIONS}

The construction of
the photometric catalog used here for M13 is described in 
\citet*{Stetson98} and \citet{Stetson00}
and is similar to that built for M5 and described in detail
in our earlier work \citep{Cohen02}.  Slitmasks
were designed focusing on the subgiants in M13 and on the region just
below the turnoff. We note that the base of the RGB in M13
is at about V = 17.7, I = 16.9, while the MSTO  is at about
V = 18.2, I = 17.8.

These slitmasks were used on the blue side 
\citep{McCarthy98} of LRIS \citep{Oke95}
at the Keck Telescope in May 2001 and in May 2002.  As the
weather was sub-optimal, four 1000 sec exposures were obtained with
the slitmask intended for the fainter stars.
The 600 g/mm grism was used with a 0.7 arcsec wide slit, yielding
a dispersion of 1.0 \AA/pixel, and a spectral resolution of 4\AA. 
The spectral coverage was from 3500 to 5500~\AA\ for essentially
all the stars.
There are 81 stars with spectra good enough for present purposes,
namely more than 450 ADU/pixel (with 2 $e^-$/ADU) in the continuum at 
4360~\AA.

Non-members are very easily recognized due to the low metallicity
of M13 and its large negative radial velocity.  Since the stars
were selected to lie on the cluster isochrone and M13 is at
high galactic latitude, one would expect very few non-members.
All of the stars in the sample presented here are members of M13.
Full details of the sample of stars and the observations
will be given in a forthcoming paper.

\section{ANALYSIS}

Our analysis essentially parallels that of \citet[hereafter BC01]{Briley01}
and the reader is referred to that work for details.

The strengths of the 4350\AA ~ CH (G) bands of our program stars were
measured via the I(CH) index of \citet{Cohen99a, Cohen99b} - a double sided index
which compares the flux removed by the G-band to the adjacent continuum.
The resulting indices (corrected for the radial velocity of M13) are plotted as a
function of I magnitude in Figure \ref{fig_allch},  where a significant spread
among both the SGB and MS stars can be seen.
As is discussed in BC01, the decrease in the spread in I(CH) near $I \approx
18$ is the result of the high temperatures and low metallicities of the M13 MS
turnoff stars (i.e., there simply is little CH formation in such stars).

In order to relate the observed indices to the underlying \cabund, we
employ a series of synthetic spectra based on MARCS \citep{Gustafsson75}
model atmospheres.
We assume that the iron abundance is constant within M13 as
suggested by the high dispersion spectroscopic studies
of low luminosity globular cluster samples by
\citet{Gratton01} and by \citet{Ramirez02, Ramirez03}.
Our MS turnoff models are those used in BC01 to which we have added an
additional 3 luminous SGB models (at \teff/\grav = 5168/2.96, 5224/3.15,
5281/3.33) and 5 fainter MS points (at 6118/4.47, 6034/4.53, 5833/4.59, 5601/4.66,
5351/4.70) taken from the 16 Gyr \fe = $-$1.48 O-enhanced isochrone
grid of \citet{Bergbusch92}.
As in BC01, we adopt a geometric distance of 7.2 kpc (plus an
additional 0.11 magnitudes in
$(m-M)_V$) and a reddening of E($B-V$) = 0.02 magnitudes
following \citet{Harris96}.

From each model and a given set of C/N/O abundances, synthetic spectra
were computed using the SSG program \citep*[][and references therein]
{Bell94} and the linelist of \citet{Tripicco95} (see BC01
for further details).
The result was then convolved with a Gaussian to match
the resolution of the observed spectra and I(CH) indices measured.
Four compositions are illustrated here ([C/Fe]/[N/Fe]/[O/Fe]): 
$-0.85/+0.7/+0.4$ and $-1.1/+1.2/-0.5$ which (as in BC01) roughly match
the observed compositions of M13's CN-weak and strong bright giants
respectively, $0.0/+0.4/+0.4,$ an arbitrary ``pre-mixing'' 
composition, and $-0.5/+1.4/0.0,$ a redistribution of the ``pre-mixing'' composition
holding the sum of C+N+O constant. 
The curves plotted in Figure \ref{fig_allch}
are I(CH) indices calculated from these synthetic spectra
where it can be seen that the small scatter in I(CH) among the MS turnoff
stars is due to the intrinsic weakness of the G-band, rather than an absence
of inhomogeneities (as pointed out by BC01).

Note that among such warm relatively metal-poor stars the resulting
indices are remarkably insensitive to all but our choice of \cabund.
This is the result of the low metallicity; CN and CO play a less
important role in the molecular equilibrium for carbon.
For example, using the 5281/3.33 model ($M_I = +1.89$) with 
$-0.5/+1.4/0.0$ as a baseline, increasing [O/Fe] by
0.4 dex decreases I(CH) by 0.006, decreasing [N/Fe] by 1.4 dex
increases I(CH) by 0.006, and increasing \ciso\ to 90 has essentially no effect.
The greatest sensitivity appears to be to microturbulence ($v_t$), as increasing
$v_t$ from 1.5 to 2.5 \kms increases I(CH) by 0.010 (an 8\% increase),
which is negligible compared to the changes resulting from different
values of \cabund.

We have applied the method of \citet{Briley90} to convert the observed
I(CH) indices to corresponding C abundances: the model isoabundance curves
were interpolated to the $M_I$ of each program star, and the observed I(CH) index
converted into \cabund\ based on the synthetic indices at that $M_I$.
As such, we are assuming that a given star's $M_I$ maps to the \teff, \grav, and
mass specified by the \citet{Bergbusch92} isochrone based on our calculation of
model luminosity/color and choice of distance modulus and reddening.
Returning to the 5281/3.33 model (the faintest model used in deriving
SGB \cabund), we note that an underestimation in \teff\ of 100K
would correspond to an error  of +0.03 in $V-I$ and a derived
\cabund\ some 0.12 dex too large at [C/Fe] $=-0.5$.
Likewise, an underestimation of \grav\ by 0.1 dex would be equivalent to a
shift of $-0.25$ in $M_I$ and  $-0.08$ dex in \cabund.
Essentially identical changes result when considering our most luminous
SGB model (5168/2.96).
Thus if one takes 100K and 0.1 dex as reasonable errors in \teff\ and \grav,
we estimate our uncertainties in \cabund\ to be roughly 0.1 dex.

\section{RESULTS}

The resulting \cabund\ values for the subgiants only
are plotted in Figure \ref{fig_chbright}, as well as the
\cabund\ values derived by \citet{Suntzeff81} and \citet{Smith96} for
more luminous cluster members.
Immediately obvious from Figure \ref{fig_chbright}
is a marked decline in \cabund\  with
increasing luminosity among the M13 giants - the mean 
 \cabund\ of Sunzteff's ``RGB'' sample  is $-0.9 \pm 0.3$ dex, while for
the present SGB stars the value is $-0.4 \pm -0.2$ dex.
This pattern in \cabund\ was noted by Suntzeff, but the depletions
appear to set in at a luminosity below the limit of his sample.
Such as result can most easily be understood in the context of the deep-mixing
framework where C-depleted material from the CN-burning region above
the H-burning shell is progressively circulated into the stellar envelope
as discussed above.
We note however that changes in the distribution of O, Na and Al in M13 are
suggested by \citet{Pilachowski96}, \citet{Kraft97}, and \citet{Cavallo00}
to occur only among its most luminous giants, i.e., $V < 13$ which is also the
point at which the C abundances drop precipitously in Figure~\ref{fig_chbright}.
It would thus appear that during the bulk of RGB ascent, mixing only into CN-cycle
regions is taking place in M13.

\begin{figure}
\plotone{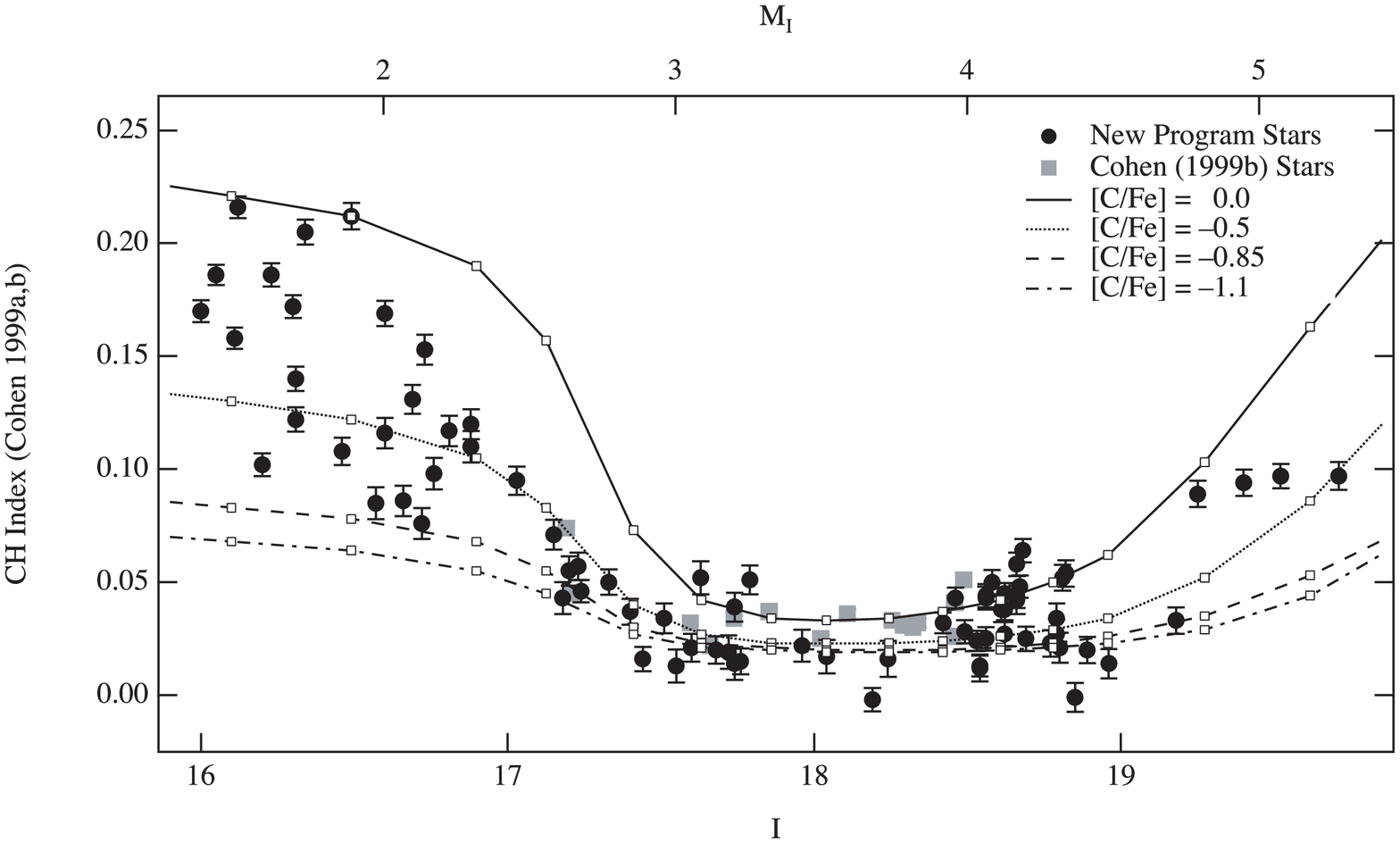}
\caption{Observed and synthetic I(CH) indices 
for stars in M13 are plotted as a function
of luminosity. Also plotted are the indices from \citet{Cohen99b}. One sigma error bars
based on Poisson statistics are included. A large spread in I(CH) is
present for both the SGB and pre-MS turnoff stars, as well as among the stars at the
base of the turnoff. Using the model isoabundance lines as a guide, 
the scatter in I(CH) appears consistent with a constant range in \cabund\ 
over the entire luminosity range.
\label{fig_allch}}
\end{figure}

\begin{figure}
\plotone{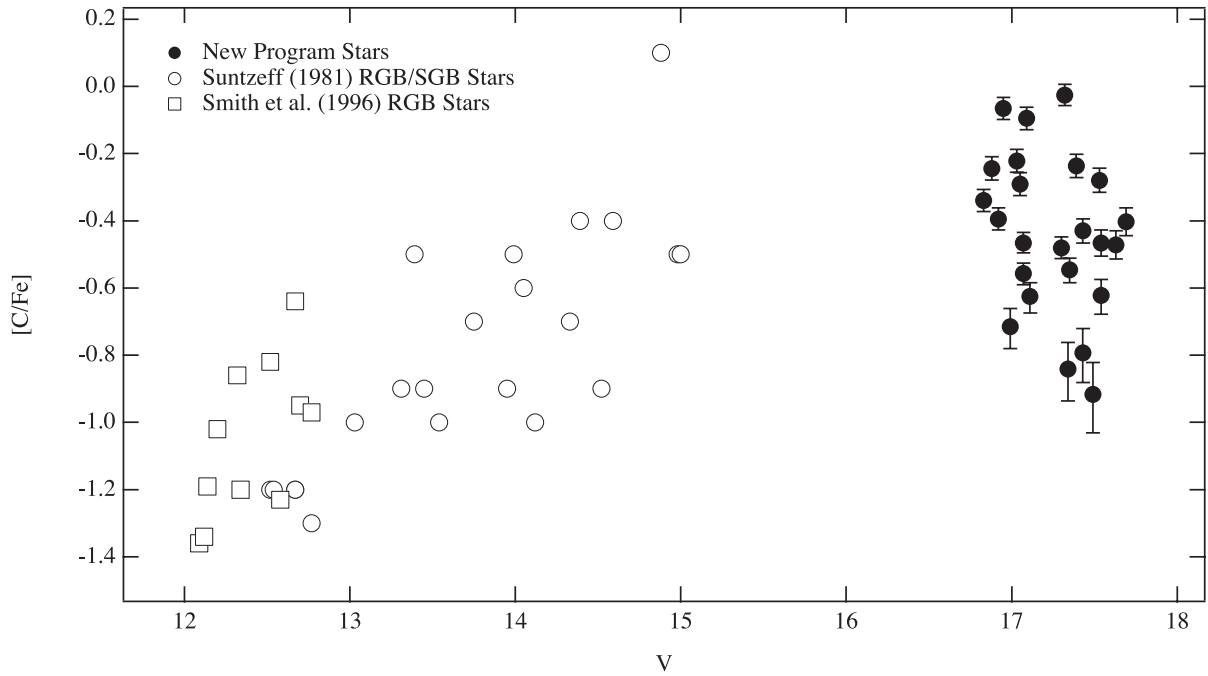}
\caption{Carbon abundances derived from the present SGB stars
in M13 are plotted with the results of \citet{Suntzeff81} and \citet{Smith96}.
Although a large and consistent scatter in \cabund\ apparently exists from the SGB
to the tip of the RGB, the decline in \cabund\ with luminosity
near the RGB tip is striking.
\label{fig_chbright}}
\end{figure}

At the same time, a large scatter in \cabund ~ (roughly 0.8 dex), comparable
to that found among the more luminous stars, is present in the LRIS SGB sample
in Figure~\ref{fig_chbright}.
Figure~\ref{fig_allch} further suggests a large dispersion in C abundance is also
present at luminosities below the MS turnoff. 
(We have no information on N abundances at present.)
These inhomogeneities are in place well below the LF bump
and the operation of proposed deep mixing mechanisms.
Thus it appears that the action of an external process has shaped much of
the star-to-star scatter present in M13, i.e., the large range of C, and likely
other light abundances found among the giants (at least fainter than $V = 13$)
is to a great extent {\it not} the result of varying mixing efficiencies
within the stars themselves. 

If we assume the initial \cabund\ values to be roughly solar in the
proto-M13 gas cloud, the C-poor stars
in our sample might be explained by the  incorporation of C-poor, N-rich AGB
ejecta.
However, the large spread in C abundances seen here requires far more than
the ``pollution'' of the stellar surfaces by simple accretion of AGB ejecta - such
surface contaminations would be quickly erased by the deepening convective
envelope following the MS phase.
This is clearly not observed and a substantial fraction of the stars' total mass
must therefore be homogeneous in these elements.
This in turn requires the accretion of  not just of a sprinkling but rather of a
significant amount of material: if we consider a
typical C-poor M13 star (\cabund\ $= -0.6$), some 70\% of the star's mass
must be captured ejecta (assuming the accreted matter to be completely
free of C).
While this initially seems implausible, we note the recent calculations of
\citet{Thoul02} which demonstrate that large accumulations
may be possible, particularly in clusters with small core radii, e.g. 47 Tuc
where they estimate as much as 80\% of the mass of a 1 M\subsun ~ star
could be accreted material.
Unfortunately, the core radius of M13 is considerably larger, and similar
calculations yield an expected accretion of only 8\%.
Whether this is a problem that can be overcome with more detailed accretion
models (e.g., Thoul et al. included several assumptions such as chaotic orbits
with stars spending  20\% of their time in the cluster core, which was also the
spatial extent of the central gas reservoir), a reflection of changes in the
structure/dynamics of M13 since its formation, or can be better explained by
an alternative mechanism remains to be seen.

The work presented here implies we are seeing the effects of a significant
external ``contamination'' of unexplained origin
upon which a poorly understood mixing mechanism acts during the
later stages of RGB ascent.
However, there remain a number of issues to address:
The critical region just fainter than the M13 LF bump at
$V \approx 14.75$ has yet to be explored, leaving
the exact luminosity of C-depletion onset uncertain.
Also, in comparing the present results with those of \citet{Suntzeff81} and
\citet{Smith96}, we are comparing abundances based on different instruments,
indices, and analysis tools - note the apparent 0.3 dex offset between the C
abundances of Suntzeff and Smith in Figure~\ref{fig_chbright}.
Furthermore, we urgently need a larger sample of stars below the MSTO in M13.
To address these items, we will be returning to M13 in the near future.

\acknowledgements
The entire Keck/LRIS user communities owes a huge debt to 
Jerry Nelson, Gerry Smith, Bev Oke, and many other 
people who have worked to make the Keck Telescope and LRIS  
a reality and to operate and maintain the Keck Observatory. 
We are grateful to the W. M.  Keck Foundation for the vision to fund
the construction of the W. M. Keck Observatory. 
We wish to express our thanks to Roger Bell whose SSG code was
instrumental in this project.
Partial support was provided by the National Science Foundation under
grant AST-0098489 to MMB and grant AST-9819614 and AST-0205951 to JGC and by
the F. John Barlow professorship (MMB).

\clearpage

\end{document}